\documentclass[12pt]{article}
\normalbaselines
\begin{document}
\title{{\bf\Large{Geometric quantization on a coset space $G/H$}}}
\author{\large{Masaomi Kimura}\\ \it{\small{Institute for Cosmic Ray
Research, University of Tokyo}}\\ \it{\small{Midoricho, Tanashi,
Tokyo 188 JAPAN}}}
\date{ }
\maketitle

\begin{abstract}

    Geometric quantization on a coset space $G/H$ is considered,
    intending to recover Mackey's inequivalent quantizations.  It is
    found that the inequivalent quantizations can be obtained by
    adopting the symplectic 2-form which leads to Wong's equation.
    The irreducible representations of $H$ which label the
    inequivalent quantizations arise from Weil's theorem, which
    ensures a Hermitian bundle over $G/H$ to exist.

\end{abstract}
\vfill
\eject

\newcommand{\be}{\begin{equation}}
\newcommand{\ee}{\end{equation}}
\newcommand{\bea}{\begin{eqnarray}}
\newcommand{\eea}{\end{eqnarray}}
\newcommand{\bdpm}{\begin{displaymath}}
\newcommand{\edpm}{\end{displaymath}}
\newcommand{\bra}[1]{\langle#1}
\newcommand{\ket}[1]{|#1\rangle}
\newcommand{\rar}{\rightarrow}
\newcommand{\HFM}{\upsilon_{q}}
\newcommand{\HFMC}{\upsilon}
\newcommand{\cots}[1]{T^{\ast}#1}
\newcommand{\LG}{\mathsf{g}}
\newcommand{\LH}{\mathsf{h}}
\newcommand{\LR}{\mathsf{r}}
\newcommand{\SEC}{\sigma_{\alpha}}
\newcommand{\SECB}{\sigma_{\beta}}
\newcommand{\WG}{\wedge}
\newcommand{\SDR}{(\SEC^{-1}d\SEC)|_{\LR}}
\newcommand{\VR}[1]{\frac{\partial}{\partial #1}}
\newcommand{\RO}[1]{\beta_{#1}}
\newcommand{\SRO}[1]{\alpha_{#1}}
\newcommand{\PRO}[1]{\varphi_{#1}}
\newcommand{\CH}{{\cal T}}
\newcommand{\secl}{\phi^{(K)}}
\newcommand{\HE}{{\bf 1}}
\newcommand{\tr}{\mbox{tr}}
\newcommand{\ses}{\phi}
\newcommand{\La}{{\bf L}} 
\newcommand{\Om}{\Omega}
\newcommand{\om}{\omega}
\newcommand{\tha}{\theta}
\newcommand{\VF}{{\cal U}}
\newcommand{\ind}{\rfloor}
\newcommand{\PB}[2]{\{#1,#2\}}
\newcommand{\CN}[1]{\nabla_{#1}}
\newcommand{\PL}{{\cal P}}
\newcommand{\hh}{{\bf H}}
\newcommand{\CONJ}[1]{{#1}^{\ast}}
\section{Introduction}\label{intro}

Geometric quantization\cite{Ko,SW,Wo,Sn} is the method of constructing 
a wave function on the phase space of a classical system and 
restricting it to be covariantly constant along a generalized 
momentum.  It is a powerful approach to quantizing on a manifold which
is topologically nontrivial.  

In this paper, we consider quantization of a classical system on a
coset space $G/H$, where $G$ is a compact Lie group and $H$ a
semisimple subgroup of $G$.  Geometric quantization has previously
been applied to coset spaces interpreted as phase spaces\cite{Hu}.  In
this study, however, a coset space $G/H$ is regarded as a
configuration space rather than a phase space, which means that the
phase space of interest is the cotangent bundle $T^{\ast}(G/H)$.

There are many approaches to quantization on a coset space, for example, a
system of imprimitivity (Mackey \cite{Mac}), a canonical group (
Isham \cite{Is}), a $C^{\ast}$-algebra (Landsman
\cite{CAST}\cite{LL}), generalized Dirac quantization (McMullan,
 Tsutsui \cite{MT}) and so forth.  One of the most important 
characteristics of quantization on $G/H$ is the existence of 
inequivalent quantizations, which were found by Mackey\cite{Mac}.  The 
inequivalent quantizations on $G/H$ are labeled by the unitary 
representations of $H$, and Hilbert spaces obtained using those 
quantizations belong to different sectors, if they do not belong to 
the same representation space of $H$.
 
The purpose of the present paper is to investigate how Mackey's
inequivalent quantizations are recovered in the context of geometric
quantization.  We construct a symplectic 2-form on $T^{\ast}(G/H)$ by
Hamiltonian reduction from a system on $T^{\ast}G$. The Lagrangian
leading to Wong's equation \cite{Wg,BMSS} is used as a guiding
principle to find the constraints which implement the reduction, on
the ground that it allows us to identify $H$ with the internal
symmetry of the system and it is also an effective Lagrangian obtained
by Mackey's quantization \cite{MT}.  After that, we construct a Hilbert
space on $G/H$ and find that Weil's condition, which ensures the
Hilbert space to exist, leads precisely to Mackey's inequivalent
quantizations. Recently, geometric quantization of
cotangent bundle with symmetry is considered in \cite{Rob1,Rob2}. Whereas
these papers put an emphasis on the inequivalent quantizations of
operators, our primary concern in this paper will be with the Hilbert
spaces pertinent to the quantizations.

This paper is organized as follows. Section 2 is devoted to a brief
review of geometric quantization.  In section 3, we consider geometric
quantization on a phase space whose configuration space is a coset
space $G/H$ with (and without) $H$ as the internal symmetry. 
The last section is devoted to the conclusion.

\section{A brief review of geometric quantization}\label{geo}

We first review the method of geometric quantization which we apply
later to $G/H$.  Geometric quantization consists of two procedures,
prequantization and a choice of polarization.  This review is mainly
based on \cite{SW,Wo}.

\vspace{5 mm}Let $M$ be a manifold with a closed 2-form $\om$.  If
$\om$ is nondegenerate, it is called a symplectic 2-form, and we can
always construct the Poisson bracket on the phase space from $\om$. 
Since $\om$ is closed, we can write locally, \be\label{e1} \om=d\tha,
\ee where $\tha$ is a 1-form called the canonical 1-form. If
$\om$ is degenerate, $M$ with $\om$ is called the presymplectic 
manifold.

For simplicity, we illastrate this approach of quantization with the
special case where $M$ is a cotangent bundle of $Q$ and $\om$ is
symplectic. $M$ is locally covered by the coordinate system
$(q^{i},p_{j})$ and $\tha$ is locally written as $p_{i}dq^{i}$.  We
construct a Hilbert space on the classical phase space $M$. Consider a
complex line bundle as a candidate of the Hilbert space.  A wave
function is defined as a section. The complex line bundle which we are
interested in has following two additional structures
\bea
\label{e2}\mbox{connection}&:&\nabla \ses =
(d-\frac{i}{\hbar}\theta) \ses\\
\label{e3}\mbox{Hermitian metric}&:&(\ses_{1},\ses_{2})
=\CONJ{\ses_{1}}\cdot \ses_{2}
\eea
for each section $\ses_{1}$ and $\ses_{2}$.
$\rfloor$ denotes an inner derivative and
$\CONJ{\ses}$ is the comprex conjugation of $\ses$. We call this line bundle
Hermitian line bundle and identify it with the Hilbert
space on $M$. This procedure is prequantization.

Weil's theorem tell us that such sections on the Hermitian line bundle
exist if the symplectic 2-form $\omega$ satisfies the condition
\be\label{e4}
\frac{1}{2\pi\hbar}\int_{S}\omega\in {\bf Z},
\ee
where $S$ is a 2-dimensional closed surface on $M$.

A physical wave function, however, must be (covariantly) constant
along momenta in the position representation, in order to make the
representation of operators irreducible.  Therefore we impose the
restriction on wave functions that the connection of the physical wave
functions with respect to vectors along (generalized) momenta must
vanish. The set of the vectors is called polarization, and this procedure is
a choice of polarization.

In order to make an inner product invariant under a coordinate
transformation(in the position representation), we introduce a
1/2-P-form $\HFM$, which is covariantly
constant along momenta, and transforms as $\HFM\mapsto
(\det m)^{-\frac{1}{2}}\HFM$ under the transformation that
$dq^{i}\mapsto m^{i}_{ j}dq^{j}$. Then we can define a wave function,
an inner product and an operator corresponding to a classical
observable $\varrho$ as
\bea
\label{e5}\psi(q)&=&\ses(q)\cdot\HFM \\
\label{e6}\langle\psi_{1},\psi_{2}\rangle
&=&\int_{Q}\CONJ{\ses}_{1}(q)\ses_{2}(q)\CONJ{\HFM}\HFM d^{n}q \\
\label{e7}\delta_{\varrho}\psi&=&
(\frac{\hbar}{i}\nabla_{\xi_{\varrho}}+\varrho)
\ses\cdot\HFM+\ses\cdot(\frac{\hbar}{i}{\cal L}_{\xi_{\varrho}})\HFM,
\eea
where $\xi_{\varrho}$ is a Hamiltonian vector field defined as
\be
d\varrho=-\xi_{\varrho}\rfloor\omega,
\ee
and $\cal{L}_{\xi_{\varrho}}$ is a Lie derivative which acts
on $\HFM$.

If $\om$ is degenerated, $M$ is reduced to the space where $\om$ is
symplectic. On $M$, there are vectors $\{\zeta\}$ which satisfies
$\zeta\rfloor\om=0$, and the reduced space is realized by identifying
points along the integral curves of $\{\zeta\}$. Thus if $S^{2}$ in
eq.(\ref{e4}) is a closed surface in the reduced space, then in $M$
it is a surface $W$ whose boundary $\gamma$ is identified in the reduced
space. This replaces eq.(\ref{e4}) with the condition \be\label{e9}
\frac{1}{2\pi\hbar}\int_{W}\om=
\frac{1}{2\pi\hbar}\oint_{\gamma}\tha\in {\bf Z}, \ee where we used
Stokes' theorem.

\section{Geometric quantization on a coset space $G/H$}\label{gqoc}

In this section, we consider geometric quantization on a coset space
$G/H$. A system we are interested in is a phase space $T^{\ast}(G/H)$
with an internal symmetry $H$, which is obtained by reduction of
$T^{\ast}G$ under certain constraints. 

We first show in subsection~\ref{ncq} that the naive set of
constraints that ``momenta'' associated with $H$ should be zero does
not lead to inequivalent quantizations. Thus we impose another set of
constraints which we find by comparing the Lagrangian of a free
particle on $G$ with the Lagrangian for Wong's equation, in subsection
~\ref{coq}, and find that this leads to inequivalent quantizations in
subsection ~\ref{coqgh}.

\subsection{Quantization under naive constraint conditions} \label{ncq}
 
Firstly, we begin by reducing $\cots{G}$ to $\cots{(G/H)}$
under the naive constraints fixing momentum associated with
 $H$ to be zero. 

The canonical 1-form on $\cots{G}$ is already given by
\be \label{c.1.1}
\Theta_{G}=\tr(\tilde{R}g^{-1}dg),\hspace{10 mm}\tilde{R}\in \LG,
\ee
where $\LG$ is the Lie algebra of $G$. $g^{-1}dg$ is a
Maurer-Cartan 1-form\cite{EGH}, and $\tilde{R}$ is its conjugate
momentum.
The Lie algebra of $G$ can be decomposed as $\LG=\{ T^{a} \}
\oplus\{ T^{i} \}$. $\{T^{i}\}$ is a basis of the Lie algebra
$\LH$ of a Lie group $H$. 
$\{T^{a}\}$ is a basis for $\LR$ being orthogonal to $\LH$. The
normalization of trace of these bases is $\tr(T^{m}T^{n})=\delta^{mn}$,
where $T^{m}, T^{n}\in\LG$.
$\tilde{R}$ can be decomposed as 
\bdpm
\tilde{R}=\tilde{R^{a}}T^{a}+\tilde{R^{i}}T^{i}.
\edpm

In order to reduce the phase space from $\cots{G}$ to $\cots{(G/H)}$,
we must restrict the degrees of freedom corresponding to $\LH$. 
Thus we do this by imposing the constraints as follows
\be\label{c.1.1.5}
 \tilde{R}^{i}\equiv 0, 
\ee
for all $i$.
This means that $\tilde{R}\in \LR$.

$\varpi_{0}:G\rar G/H$ is regarded as a bundle on $G/H$. Let 
$\{ U_{\alpha}\}$ be a set of open coverings of $G/H$. Thus $g\in G$ in
eq.(\ref{c.1.1}) can be written as 
\be\label{c.1.1.6}
g=\SEC (q)h,
\ee
where $q=\varpi_{0}(g)\in G/H$, $h\in H$, and  $\SEC$ is a section such that
$\SEC:U_{\alpha}\rar G$.  Using these $q$ and $h$, we can rewrite
eq.(\ref{c.1.1}) in the form
\be \label{c.1.2}
\Theta_{G}=\tr(\tilde{R}h^{-1}dh)+\tr(\tilde{R}h^{-1}(\SEC^{-1}d\SEC)h).
\ee
Since  $\tilde{R}\in \LR$, the first term on the right hand side
vanishes. Define as $R=h\tilde{R}h^{-1} \in \LR$, and
\bea \label{c.1.3}
\Theta = \Theta_{G}{|}_{\tilde{R}\in \LR} = \tr(R(\SEC^{-1}d\SEC)|_{\LR}), 
\eea
where the sign $|_{\LR}$ denotes the restiction to $\LR$.
The symplectic 2-form $\omega$ is defined as
\bea \label{c.1.4}
\omega &=& d\Theta \nonumber \\
&=& \tr(dR\WG \SDR)+\tr(R d\SDR) .
\eea

In fact, this symplectic 2-form is exact, since the canonical 1-form
$\Theta$ is defined globally. This can be seen from that $\Theta$ is
invariant under the gauge transformation $\SEC=\sigma_{\beta}h_{\beta
\alpha}$ on $U_{\alpha}\cap U_{\beta}$, where $\SDR$ and $R$ transforms
as $\SDR\rar (h_{\beta \alpha}^{-1}\SECB^{-1}d\SECB h_{\beta
\alpha})|_{\LR}$ and $R\rar (h_{\beta \alpha}^{-1}Rh_{\beta
\alpha})|_{\LR}$.

Secondly, we consider prequantization.
The quantization condition on $\cots{(G/H)}$ is satisfied, because
\be \label{c.1.5}
\frac{1}{2\pi \hbar}\int_{S}\omega=\frac{1}{2\pi
\hbar}\int_{S}d\Theta=0\in{\bf Z},
\ee
where $S$ is a 2-dimensional closed surface on $\cots{(G/H)}$.

Thus we can define a Hermitian line bundle $L$ on $\cots{(G/H)}$.  Let
$\Gamma(L)$ denote a set of sections on $L$. The connection for
$\Gamma(L)$ is defined as
\be \label{c.1.6}
\nabla\phi=d\phi-\frac{i}{\hbar}\Theta\phi,
\ee
for $\phi(R,q)\in\Gamma(L)$. 
The Hermitian metric compatible with this connection is 
\be \label{c.1.7}
(\phi_{1},\phi_{2})(R,q)=\CONJ{\phi_{1}}(R,q)\phi_{2}(R,q),
\ee
where $\phi_{1},\phi_{2}\in \Gamma(L)$.

Lastly, we choose a polarization as $\{ \VR{R_{a}}\}$. Thus
\be \label{c.1.8}
\nabla_{\VR{R_{a}}}\phi=\VR{R_{a}}\rfloor d\phi=\VR{R_{a}}\phi=0.
\ee
Therefore, $\phi$ depends only on $q$, and can be written as $\phi(q)$.
Let $\HFM(\{ \VR{R_{a}}\})$ denotes the 1/2-P-form. The wave function is 
written as $\phi(q)\HFM$.
The inner product of wave functions is defined as
$\int_{G/H}(\phi_{1},\phi_{2})(q)\CONJ{\HFM}\HFM d^{n}q$.

Inequivalent quantizations do not emerge under the constraint
eq.(\ref{c.1.1.5}), as can be seen from eq.(\ref{c.1.5}).  This is a
situation similar to the case of Dirac's naive constraints in
\cite{MT}. Now we discuss what constraints
should be imposed on $T^{*}G$ to obtain symplectic 2-form on
$T^{*}(G/H)$ which correspond to inequivalent quantizations.

\subsection{Derivation of the symplectic 2-form on $G/H$ leading to 
inequivalent quantizations} \label{coq}
We start by decomposing the degrees of freedom of a system on $G$ 
to $G/H$ and $H$, and investigate what constraints should be imposed.

The first order Lagrangian of a free particle on $G$ is
\be \label{c.2.1}
\La=\tr(\tilde{R}g^{-1}\dot{g})-\frac{1}{2}\tr(\tilde{R}^{2}). 
\ee
One can see that this lead to the Lagrangian
$\La=\frac{1}{2}\tr(g^{-1}\dot{g})^{2}$ by eliminating $\tilde{R}$,
using equations of motion for $\tilde{R}$.

We rewrite $g^{-1}dg$, using eq.(\ref{c.1.1.6}), as 
\bea \label{c.2.3}
g^{-1}dg&=&h^{-1}{\SEC}^{-1}d\SEC h+h^{-1}dh \nonumber \\
&=&(h^{-1}{\SEC}^{-1}d\SEC h+h^{-1}dh){|}_{\LH}
+(h^{-1}{\SEC}^{-1}d\SEC h){|}_{\LR},
\eea
where $|_{\LH}$ denotes the restriction to $\LH$.
Here we define $A = ({\SEC}^{-1}d\SEC){|}_{\LH}$, called
H-connection\cite{LL,MT}, because,
under the gauge transformation as $\SEC\mapsto\SEC h$,
$({\SEC}^{-1}d\SEC){|}_{\LH}$ transforms as a vector potential
of non Abelian gauge field whose gauge group is $H$.
Using eq.(\ref{c.2.3}), we can rewrite eq.(\ref{c.2.1}) as
\bea \label{c.2.5}
\La&=&\tr(\tilde{R}{|}_{\LR}(h^{-1}\SEC^{-1}\dot{\SEC}h){|}_{\LR})
-\frac{1}{2}\tr((\tilde{R}{|}_{\LR})^{2}) \nonumber \\ 
&\hspace{1 cm}&+\tr(\tilde{R}{|}_{\LH}h^{-1}\dot{h})+
\tr(h\tilde{R}{|}_{\LH}h^{-1}\dot{A})
-\frac{1}{2}\tr((\tilde{R}{|}_{\LH})^{2}),
\eea
where we use
$\dot{A}=A_{a}\dot{q}^{a}=({\SEC}^{-1}\dot{\SEC}){|}_{\LH}$.

By using the equations of motion for $\tilde{R}_{a}$
\be \label{c.2.6}
\tilde{R}_{a}=(h^{-1}{\SEC}^{-1}\dot{\SEC}h)_{a},
\ee
we substitute this for $\tilde{R}_{a}$ in eq.(\ref{c.2.5}),
and eliminate $\tilde{R}_{a}$. Thus the Lagrangian (\ref{c.2.5}) becomes
\be \label{c.2.7}
\La=\frac{1}{2}\tr((\SEC^{-1}\dot{\SEC}{|}_{\LR})^{2})
+\tr(\tilde{R}{|}_{\LH}h^{-1}\dot{h})+\tr(h\tilde{R}{|}_{\LH}h^{-1}\dot{A})
-\frac{1}{2}\tr((\tilde{R}{|}_{\LH})^{2}), 
\ee
where $\SEC^{-1}\dot{\SEC}{|}_{\LR}=
\SEC^{-1}\partial_{a}\SEC{|}_{\LR}\dot{q}^{a}$, and the metric on
$G/H$ is given by $g_{ab}=\tr(\sigma^{-1}\partial_{a}\sigma{|}_{\LR}
\sigma^{-1}\partial_{b}\sigma{|}_{\LR})$.

Now we recall Wong's equation is known to be an effective lagrangian
which leads to Mackey's inequivalent quantizations\cite{MT} and it
describes a system coupling with non Abelian gauge
field\cite{Wg,BMSS}. Here we make use of the Lagrangian for Wong's
equation to identify $H$ as the gauge group.
The Lagrangian is
\be \label{c.2.8} \La=\frac{1}{2}g_{ab}\dot{q}^{a}\dot{q}^{b}
+i\hbar \tr(Kh^{-1}\dot{h})+i\hbar \tr(hKh^{-1}\dot{A}), 
\ee 
where $K$ is a constant element in $\LH$.

Obviously, eq.(\ref{c.2.7}) and eq.(\ref{c.2.8}) have the same form
except for $\tilde{R}{|}_{\LH}$ and $K$ \cite{MT}. By identifying these two
equations, therefore, it is possible to recognize that eq.(\ref{c.2.7})
describes quantum theory whose classical configuration space is $G/H$,
and whose internal symmetry is $H$.
Consequently, a constraint which we should impose is
\be \label{c.2.9}
\tilde{R}{|}_{\LH}\equiv i\hbar K.
\ee

The basis of $\LH$ is defined so that its Cartan subalgebra commutes with
$K$. We neglect the term
$-\frac{1}{2}\tr((\tilde{R}{|}_{\LH})^{2})$ in eq.(\ref{c.2.7}),
because it is constant under the constraints, eq.(\ref{c.2.9}).

Under the constraints, the canonical 1-form on $T^{*}G$ in eq.(\ref{c.1.1})
is reduced on $\cots{(G/H)}$ to 
\be \label{c.2.10}
\Theta=R_{a}(\SEC^{-1}d\SEC)^{a}+i\hbar \tr(hKh^{-1}A)+i\hbar \tr(Kh^{-1}dh).
\ee
The closed 2-form is
\bea \label{c.2.11}
\Omega&=&d\Theta \nonumber\\
&=&dR_{a}\WG(\SEC^{-1}d\SEC)^{a}+R_{a}d(\SEC^{-1}d\SEC)^{a}
+i\hbar \tr(d(hKh^{-1})\WG A)\nonumber \\
&\hspace{5 mm}&+i\hbar \tr(hKh^{-1}dA)-i\hbar \tr(Kh^{-1}dh\WG h^{-1}dh).
\eea

Let $\Delta$ be a set of roots of
$\LH$. $\RO{1},\RO{2}\ldots\in\Delta$ denote roots of $\LH$. With
respect to the roots, we
can decompose (complexified) $\LH$ as
\be \label{c.2.12}
\LH = \CH \oplus \LH_{\RO{1}} \oplus \LH_{\RO{2}} \oplus \ldots ,
\ee
where $\CH$ is a space spanned by the Cartan subalgebra of $\LH$, and
$\LH_{\RO{i}}$ is an eigenspace of $\CH$ and a space whose elements
has $\RO{i}$ as a root\cite{MT,Gi}.
Let $E_{\RO{i}}\in \LH_{\RO{i}}$, $E_{\RO{j}}\in \LH_{\RO{j}}$, then 
\be\label{c.2.13} 
[E_{\RO{i}},E_{\RO{j}}]
\left\{\begin{array}{ll} 
\in \LH_{\RO{i}+\RO{j}}&\mbox{if $\RO{i}+\RO{j}\in \Delta$} ,\\ 
=\frac{2(\RO{i})_{k}}{|\RO{i}|^{2}}H_{k}=H_{\RO{i}}&\mbox{if
$\RO{i}=-\RO{j}$},\\ 
=0 &\mbox{otherwise} ,\\ 
\end{array}
\right.
\ee
where $\{H_{i}\}$ is an orthonormal basis of $\CH$.
Since $h^{-1}dh\in\LH$, $h^{-1}dh$ can be expanded as
\be\label{c.2.14}
h^{-1}dh=\sum_{\SRO{j}\in\Delta_{s}}b_{j}H_{\SRO{j}}
+\sum_{\PRO{j}\in\Delta^{+}}
(B_{\PRO{j}}E_{\PRO{j}}+B_{-\PRO{j}}E_{-\PRO{j}}), 
\ee 
where $\Delta_{s}$ is a set of simple roots and $\Delta^{+}$ is a set
of positive roots. And
$B_{-\PRO{j}}=-\CONJ{B}_{\PRO{j}}$, because $h^{-1}dh$ is
anti-Hermite and $E_{\PRO{i}}^{\dagger}=E_{-\PRO{i}}$.
($\CONJ{B}$ denotes the complex conjugation of $B$.)
We substitute this for $h^{-1}dh$ in eq.(\ref{c.2.11}).
Since $K=K^{i}H_{i}\in\CH$, we find
\bea \label{c.2.15}
\lefteqn{\tr(K h^{-1}dh\WG h^{-1}dh)}\nonumber \\
&=&\hspace{-3 mm}
\tr(K(b_{i}H_{\SRO{i}}+B_{\PRO{i}}E_{\PRO{i}}+B_{-\PRO{i}}E_{-\PRO{i}})\WG
(b_{j}H_{\SRO{j}}+B_{\PRO{j}}E_{\PRO{j}}+B_{-\PRO{j}}E_{-\PRO{j}}))\nonumber \\
&=&\hspace{-3 mm}
\tr(K(\frac{1}{2}[H_{\SRO{i}},H_{\SRO{k}}]b_{i}\WG b_{k}
+[H_{\SRO{i}},E_{\PRO{k}}]b_{i}\WG B_{\PRO{k}} \nonumber \\
&\hspace{1 cm}&+[H_{\SRO{i}},E_{-\PRO{k}}]b_{i}\WG B_{-\PRO{k}}
+\frac{1}{2}[E_{\PRO{k}},E_{\PRO{l}}]B_{\PRO{k}}\WG B_{\PRO{l}}\nonumber \\
&\hspace{1 cm}&+\frac{1}{2}[E_{-\PRO{k}},E_{-\PRO{l}}]B_{-\PRO{k}}\WG
B_{-\PRO{l}}
+[E_{\PRO{k}},E_{-\PRO{l}}]B_{\PRO{k}}\WG B_{-\PRO{l}}))\nonumber \\
&=&\hspace{-3 mm}
-\sum_{\PRO{j}\in\Delta^{+}}
K_{\PRO{j}}B_{\PRO{j}}\WG \CONJ{B}_{\PRO{j}}, 
\eea
where
$K_{\PRO{j}}=\tr(KH_{\PRO{j}})=\frac{2K^{i}(\PRO{j})_{i}}{|\PRO{j}|^{2}}$.
Thus, the closed 2-form is
\bea \label{c.2.16}
\Omega&=&dR_{a}\WG(\SEC^{-1}d\SEC)^{a}+R_{a}d(\SEC^{-1}d\SEC)^{a}
+i\hbar \tr(d(hKh^{-1})\WG A)\nonumber \\
&\hspace{1 cm}&+i\hbar \tr(hKh^{-1}dA)
+i\hbar \sum_{\PRO{j}\in\Delta^{+}}
K_{\PRO{j}}B_{\PRO{j}}\WG \CONJ{B}_{\PRO{j}}. 
\eea

In fact, this 2-form is still not symplectic because it is degenerated. 
To see that, we first define duals of $b_{i}$ and $B_{\PRO{j}}$ 
as $y_{j}$ and $Y_{\PRO{j}}$, respectively,
$\langle b_{i},y_{j}\rangle=\delta_{ij}$, 
$\langle B_{\PRO{i}},Y_{\PRO{j}}\rangle=\delta_{\PRO{i},\PRO{j}}$.
In particular, we examine the term $d(hKh^{-1})$ in eq.(\ref{c.2.16}),
and it can be rewrote as
\bea \label{c.2.17}
d(hKh^{-1})&=&dhKh^{-1}-hKh^{-1}dhh^{-1} \nonumber \\
&=&h(h^{-1}dh)Kh^{-1}-hK(h^{-1}dh)h^{-1}.
\eea
And we can immediately find that the Cartan subalgebra component of 
$h^{-1}dh$ vanishes in eq.(\ref{c.2.17}), because
\be \label{c.2.18}
h(b_{i}H_{\SRO{i}})Kh^{-1}-hK(b_{i}H_{\SRO{i}})h^{-1}=0,
\ee
where we use that $b_{i}H_{\SRO{i}}$ and $K$ commute. Hence,
$y_{i}\rfloor d(h^{-1}Kh)=0,$ and thus, $y_{j}\rfloor \Omega=0.$
In contrast, $\VR{R_{a}}\rfloor\Omega$, $Z_{a}\rfloor\Omega$,
$Y_{\PRO{i}}\rfloor\Omega$, $\CONJ{Y}_{\PRO{i}}\rfloor\Omega$ are
nonzero, where $Z_{b}$ is defined by
$\langle (\SEC^{-1}d\SEC)_{a},Z_{b}\rangle=\delta_{ab}$.  
Thus the condition that a Hermitian line 
bundle exists is not eq.(\ref{e4}) but eq.(\ref{e9}).

Let ${\cal K}_{y}$ be an integral manifold of
$\{y_{i}\}$\cite{Wo}, which is a manifold filled with integral
curves induced by $\{y_{i}\}$. 
In order to define a well-defined Hermitian line bundle on
$\cots{(G/H)}$ with internal symmetry $H$, the condition
\be \label{c.2.21}
\frac{1}{2\pi\hbar}\oint_{\gamma}\Theta
=\frac{1}{2\pi\hbar}\oint_{\gamma}\tr(i\hbar Kh^{-1}dh)\in \bf{Z}.
\ee
must hold for all one-dimensional closed curves
$\gamma\subset{\cal K}_{y}$, as we mentioned in section\ref{geo}.
Since $h$ on $\gamma\subset{\cal K}_{y}$ is parametrized as 
$e^{-i\theta^{i}(t)H_{\SRO{i}}}$ where $\theta^{i}(T)-\theta^{i}(0)=2\pi
n^{i}  (n^{i}\in\bf{Z})$, we find
\be \label{c.2.22}
\frac{1}{2\pi\hbar}\oint_{\gamma}\Theta
=\frac{1}{2\pi}\int_{\theta^{i}(0)}^{\theta^{i}(T)}
K_{\SRO{i}}d{\theta}^{i}=K_{\SRO{i}}n^{i}.
\ee
Thus\footnote[1]{This is
suggested in \cite{Rob1}.}, for $\forall i=1,2,\ldots
\mbox{dim}H$,
\be \label{c.2.23}
K_{\SRO{i}}\in \bf{Z}.
\ee
The $K_{\SRO{i}}$'s label the inequivalent quantum theories of this system. 

Here we notice that the condition for Mackey's inequivalent
quantizations \cite{MT} is obtained by Weil's condition, eq.(\ref{c.2.21}),
in the context of geometric quantization.

\subsection{Quantization on $G/H$} \label{coqgh}

We have obtained the symplectic 2-form which corresponds to 
Mackey's inequivalent quantizations, and let us apply geometric
quantization procedure to this system.

We start with prequantization.  Let $S_{K}$ be the subgroup of
$H$ which leaves $K$ invariant under the coadjoint action such that
$s^{-1}Ks = K$ for all $s\in S_{K}$.  $S_{K}$ can be identified with
${\cal K}_{y}$, because an element of $S_{K}$ is obtained by the
exponential map of elements of Cartan subalgebra ${\cal T}$.  The
symplectic 2-form $\Omega$ defined in eq.(\ref{c.2.16}) is invariant
under gauge transformation associated with $S_{K}$.  Therefore,
$\cots{(G/H)}\times (H/S_{K})$ is the classical phase space to be
quantized.

If $K$ satisfies the condition of eq.(\ref{c.2.23}), we can 
define a line bundle  $\varpi:L\rar\cots{(G/H)}\times (H/S_{K})$.
Let $\secl(R,q,h)$ be a section on $L$.
From the above argument, $\secl(R,q,h)$ should be invariant under
the right action of $s\in S_{K}$ up to a phase factor.
In other words, $\secl(R,q,h)$ should be constant
under the parallel translation in the direction of $\{y_{i}\}$,
because we identify all points in ${\cal K}_{y}$.
The Hermitian metric and connection for sections are given as
\bea \label{c.2.24}
(\secl_{1},\secl_{2})(R,q,h)=\CONJ{\secl_{1}}(R,q,h) \secl_{2}(R,q,h)\\
\label{c.2.25}
\nabla\secl(R,q,h)=(d-\frac{i}{\hbar}\Theta)\secl(R,q,h),
\eea
respectively.
Thus $\secl(R,q,h)$ satisfies the condition $\nabla_{y_{i}}\secl(R,q,h)=0$,
and we obtain its transformation under the action of
$s=e^{-i\theta^{i}H_{\SRO{i}}}\in S_{K}$ as
\bea \label{c.2.23.5}
\secl(R,q,hs)&=&\exp(-\int_{e}^{s}\tr(K(h^{-1}dh)))\secl(R,q,h)
\nonumber \\
&=&\exp(iK_{\SRO{i}}\theta^{i})\secl(R,q,h).
\eea

Secondly, we choose a polarization.
$\Omega$ which is defined in eq.(\ref{c.2.16}) shows
us that it is possible to choose a polarization
as the set $\{\VR{R_{a}},\CONJ{Y}_{\PRO{i}}\}$.
Thus, conditions which a quantizable section fulfills is that
\bea 
\label{c.2.27.1}\nabla_{\VR{R_{a}}}\secl(R,q,h)=0 ;\\
\label{c.2.27.2}\nabla_{\CONJ{Y}_{\PRO{i}}}\secl(R,q,h)=0 .\\
\nonumber 
\eea
Eq.(\ref{c.2.27.1}) shows that the section $\secl$ does not depend on
$R_{a}$'s. Thus we let $\secl(q,h)$ represent $\secl$ as the solution
of eq.(\ref{c.2.27.1}).

Let us consider the condition eq.(\ref{c.2.27.2}). Though the solution of the 
equation is given in \cite{MT}, we explain the argument done there for
completeness.
Recall the orthonormal condition of an irreducible unitary representation,
$\pi^{\chi}_{\mu\nu}(h)$, of $H$ 
\be \label{c.2.28}
d_{\chi}\int_{H}d\mu(h) \pi^{\chi}_{\mu \nu}(h)\pi^{\chi'\ast}_{\rho\sigma}(h)
=\delta_{\mu \rho}\delta_{\nu \sigma}\delta^{\chi \chi'}V_{H} ,
\ee
where $\pi^{\chi\ast}_{\mu \nu}$ represents a complex conjugation
of $\pi^{\chi}_{\mu \nu}$.
$d_{\chi}$ represents the dimension of the representation whose highest
weight is $\chi$ and $V_{H}$ the volume of $H$ $(=\int_{H}d\mu(h))$,
respectively, and $d\mu(h)$ is a Haar measure of $H$.
Since $\pi^{\chi}_{\mu \nu}(h)$
spans a complete set, one can expand the section as
\be \label{c.2.29}
\secl(q,h)=\sum_{\chi,\mu,\nu}\tilde{\phi}^{\chi}_{\mu \nu}(q)
\pi^{\chi \ast}_{\mu \nu}(h) .
\ee
Note that we can interpret $\pi^{\chi}_{\mu \nu}(h)$ as
\be \label{c.2.30}
\pi^{\chi}_{\mu \nu}(h)=\langle \chi ,\mu |\pi^{\chi}(h)|\chi ,\nu
\rangle ,
\ee
where $|\chi ,\mu\rangle$ is an eigenstate of the Cartan subalgebra
of $\LH$: $H_{\RO{i}}|\chi ,\mu\rangle = \mu_{(\RO{i})}|\chi
,\mu\rangle$, where $\mu_{(\RO{i})}=\frac{2\RO{i}\cdot\mu}{|\RO{i}|^{2}}$.
$Y_{\PRO{i}}$ is the dual of $B_{\PRO{i}}=\tr(h^{-1}dh E_{-\PRO{i}})$.
The explicit form of $Y_{\PRO{i}}$ is 
\be \label{c.2.31}
Y_{\PRO{i}}=C\cdot \tr(hE_{\PRO{i}}\VR{h^{T}}),
\ee
where $C=\{\tr(E_{\PRO{i}}E_{-\PRO{i}})\}^{-1}$.
Eq.(\ref{c.2.31}) indicates that $Y_{\PRO{i}}$ is a left-invariant
vector field. Thus,
\bea \label{c.2.32}
\CONJ{Y}_{\PRO{i}}\pi^{\chi\ast}_{\mu \nu}(h)
&=&C\cdot {\langle \chi ,\mu |\pi^{\chi}(hE_{\PRO{i}})
|\chi ,\nu \rangle}^{\ast}
\nonumber \\
&=&C\cdot
{\langle\chi,\mu|\pi^{\chi}(h)\pi^{\chi}(E_{\PRO{i}})|\chi,\nu\rangle}^{\ast}
\eea
This is equal to zero if $\nu=\chi$.
Thus, owing to eq.(\ref{c.2.29}), $\secl(q,h)$ which fulfills the
condition eq.(\ref{c.2.27.2}), is 
\be \label{c.2.33}
\secl(q,h)=\sum_{\chi,\mu}\tilde{\phi}^{\chi}_{\mu \chi}(q)
\pi^{\chi \ast}_{\mu \chi}(h).
\ee
For highest weight, we use the notation 
$H_{\SRO{i}}|\chi,\chi\rangle = \chi_{(\SRO{i})}|\chi,\chi\rangle $,
and $\chi_{(\SRO{i})} \in {\bf Z}$.
Since the equation that $\pi^{\chi\ast}_{\mu
\chi}(s)=\delta_{\mu\chi}\exp(i\chi_{(\SRO{i})}\theta^{i})$ holds
for $s\in S_{K}$ defined in eq.(\ref{c.2.23.5}),
\be \label{c.2.34}
\secl(q,hs)=\sum_{\chi,\mu}\tilde{\phi}^{\chi}_{\mu \chi}(q)
\pi^{\chi\ast}_{\mu \chi}(h)\exp(i\chi_{(\SRO{i})}\theta^{i}).
\ee
Thus, we find that the section which we want is the component of 
$\chi_{(\SRO{i})}=K_{\SRO{i}}$,
comparing eq.(\ref{c.2.34})  with eq.(\ref{c.2.23.5}).

A wave function on $G/H$ with an internal symmetry $H$ is the form that
\be \label{c.2.35}
\secl(q,h)\HFMC=\sum_{\mu}\psi_{\mu}(q) \pi^{\chi\ast}_{\mu \chi}(h)\HFMC,
\ee
where $\HFMC$ is a 1/2-P-form on $G/H$.

The inner product of the wave functions is
\bea \label{c.2.36}
\langle \secl_{1},\secl_{2}\rangle&=&\int_{G/H}\CONJ{\HFMC}\HFMC d^{n}q
\left(\frac{d_{\chi}}{V_{H}}\right) \int_{H}d\mu(h)
(\secl_{1}, \secl_{2})(q,h)\nonumber \\
&=&\int_{G/H}d^{n}q\sum_{\mu}\psi_{1\mu}^{\dagger}(q)\psi_{2\mu}(q)
\CONJ{\HFMC}\HFMC,
\eea
where we use eq.(\ref{c.2.28}) and eq.(\ref{c.2.35}) for the last
equality. 

The operator corresponding to a classical observable $\varrho$ is
\be \label{c.2.36.5}
\hat{\delta}_{\varrho}=\frac{\hbar}{i}\nabla_{\xi_{\varrho}}+\varrho
+\frac{\hbar}{i}{\cal L}_{\xi_{\varrho}}.
\ee

Note that all physical information of the wave function is
contained in $\psi_{\mu}(q)$ as can be easily seen  from
eq.(\ref{c.2.36}). We can solve eq.(\ref{c.2.35}) with respect to
$\psi_{\mu}(q)$ using eq.(\ref{c.2.28}) as
\be\label{c.2.40}
\psi_{\mu}(q)\HFMC
=\frac{d_{\chi}}{V_{H}}\int_{H}d\mu(h) \secl(q,h)\pi^{\chi}_{\mu
\chi}(h)\HFMC.
\ee
Thus, $\psi_{\mu}(q)\HFMC$ is regarded as a physical wave function and
is labeled by the character of the representations
of $H$. This shows that our method by geometric quantization
reproduces Mackey's inequivalent quantizations \cite{Mac} and is
consistent with  other approaches \cite{CAST,Is,MT} as well.

\section{Conclusion}

In this paper we have quantized a classical system on a coset space
$G/H$ based on the method of geometric quantization.

We constructed the classical system on $T^{\ast}(G/H)$ by Hamiltonian
reduction from $T^{\ast}G$.  The naive set of constraints in
eq.(\ref{c.1.1.5}) which implements the reduction was found to lead
only to the trivial sector of inequivalent quantizations.  However,
the comparison of the Lagrangian for a free particle on $G$ with the
one that leads to Wong's equation yields the guiding principle for
finding the symplectic 2-form leading to Mackey's inequivalent
quantizations.

The important point is that the inequivalent quantizations derive from
Weil's theorem applied to presymplectic manifolds.  In contrast to
\cite{Rob1,Rob2} which characterized the superselection sectors by
operators, we did so by Hilbert spaces. This characterization arises
from the procedure of restricting wave functions to those which are
covariantly constant along the integral curves of the polarization.

\vspace{10 mm}{\Large{\bf Acknowledgement}}

\vspace{5 mm}
The author would like to thank Dr. Izumi Tsutsui and Dr. Shogo
Tanimura for advice.

\end{document}